%% file: main.tex
\documentclass[conference]{IEEEtran}
\IEEEoverridecommandlockouts
\usepackage{cite}
\usepackage{amsmath,amssymb,amsfonts}
\usepackage{algorithmic}
\usepackage{graphicx}
\usepackage{textcomp}
\usepackage{xcolor}
\usepackage{siunitx}
\usepackage[acronym]{glossaries}
\usepackage{listings}
\usepackage{cleveref}
\usepackage{adjustbox}
\usepackage{array}
\usepackage[pscoord]{eso-pic}
\usepackage[caption=false]{subfig}

\def\BibTeX{{\rm B\kern-.05em{\sc i\kern-.025em b}\kern-.08em
    T\kern-.1667em\lower.7ex\hbox{E}\kern-.125emX}}

\begin{document}

\def\notodo{}

\def\tablermws{0cm}

\ifx\blind\undefined
    \def\wheretofind{\texttt{https://github.com/pulp-platform/iDMA}}
\else
    \def\wheretofind{\todo{[URL omitted for blind review]}}
\fi

\ifx\notodo\undefined
    \newcommand\todo[1]{\textit{\textcolor{red}{#1}}}
\else
    \newcommand\todo[1]{#1}
\fi

 \renewcommand{\baselinestretch}{1.00}

\title{A Direct Memory Access Controller (DMAC) for Irregular Data Transfers on RISC-V Linux Systems}

\ifx\blind\undefined
    \author{%
        \IEEEauthorblockN{%
            Thomas Benz\IEEEauthorrefmark{3}\IEEEauthorrefmark{1}, %
            Axel Vanoni\IEEEauthorrefmark{3}\IEEEauthorrefmark{1}, %
            Michael Rogenmoser\IEEEauthorrefmark{1}, %
            Luca Benini\IEEEauthorrefmark{1}\IEEEauthorrefmark{2} %
        \thanks{%
            \IEEEauthorrefmark{3} Both authors contributed equally to this research.
        }
        }%
        \IEEEauthorblockA{\IEEEauthorrefmark{1}\textit{ETH Zurich}, Zurich, Switzerland}
        \IEEEauthorblockA{\IEEEauthorrefmark{2}\textit{University of Bologna}, Bologna, Italy}
        \IEEEauthorblockA{\{tbenz,axvanoni,michaero,lbenini\}@ethz.ch}
    }
\else
    \author{\Large\textit{Authors omitted for blind review}\vspace{1.5cm}\normalsize}
\fi

\maketitle

\ifx\showrevision\undefined
\else
\AddToShipoutPictureFG{%
    \put(%
        8mm,%
        \paperheight-3cm%
        ){\vtop{{\null}\makebox[0pt][c]{%
            \rotatebox[origin=c]{90}{%
                \Huge\textcolor{purple}{v\reviewpass}%
            }%
        }}%
    }%
}
\fi

\newacronym{dma}{DMA}{direct memory access}
\newacronym[firstplural=central processing units (CPUs)]{cpu}{CPU}{central processing unit}
\newacronym[firstplural=hardware abstraction layers (HALs)]{hal}{HAL}{hardware abstraction layer}
\newacronym{amba}{AMBA}{Advanced Microcontroller Bus Architecture}
\newacronym{axi}{AXI}{Advanced eXtensible Interface}
\newacronym{plic}{PLIC}{Platform-Level Interrupt Controller}
\newacronym{irq}{IRQ}{interrupt request}
\newacronym{dmac}{DMAC}{direct memory access controller}
\newacronym{fifo}{FIFO}{first in, first out}
\newacronym{ooc}{OOC}{out of context}
\newacronym{fpga}{FPGA}{field programmable gate array}
\newacronym{ml}{ML}{machine learning}
\newacronym{ar}{AR}{read address}
\newacronym{r}{R}{read data}
\newacronym{aw}{AW}{write address}
\newacronym{w}{W}{write data}
\newacronym{b}{B}{write response}
\newacronym{mcu}{MCU}{microcontroller unit}
\newacronym[firstplural=systems on chip (SoCs)]{soc}{SoC}{system on chip}
\newacronym{csr}{CSR}{configuration and status register}
\newacronym{noc}{NoC}{network on chip}
\newacronym{lut}{LUT}{lookup table}
\newacronym{ff}{FF}{flip-flop}
\newacronym{api}{API}{application programming interface}
\newacronym{ip}{IP}{intellectual property}

\begin{abstract}
With the ever-growing heterogeneity in computing systems, driven by modern machine learning applications, pressure is increasing on memory systems to handle arbitrary and more demanding transfers efficiently. %
Descriptor-based direct memory access controllers (DMACs) allow such transfers to be executed by decoupling memory transfers from processing units. %
Classical descriptor-based DMACs are inefficient when handling arbitrary transfers of small unit sizes. %
Excessive descriptor size and the serialized nature of processing descriptors employed by the DMAC lead to large static overheads when setting up transfers. %
To tackle this inefficiency, we propose a descriptor-based DMAC optimized to efficiently handle arbitrary transfers of small unit sizes. %
We implement a lightweight descriptor format in an AXI4-based DMAC. %
We further increase performance by implementing a low-overhead speculative descriptor prefetching scheme without additional latency penalties in the case of a misprediction. %
Our DMAC is integrated into a 64-bit Linux-capable RISC-V SoC and emulated on a Kintex FPGA to evaluate its performance.
Compared to an off-the-shelf descriptor-based DMAC IP, we achieve \todo{1.66$\times$} less latency launching transfers, increase bus utilization up to \todo{2.5$\times$} in an ideal memory system with \todo{64-byte-length} transfers while requiring \todo{11\%} fewer lookup tables, \todo{23\%} fewer flip-flops, and no block RAMs. %
We can extend our lead in bus utilization to \todo{3.6$\times$} with \todo{64-byte-length} transfers in deep memory systems. %
We synthesized our DMAC in {GlobalFoundries'} {GF12LP+} node, achieving a clock frequency of over \todo{1.44~GHz} while occupying only \todo{49.5~kGE}.
\end{abstract}


\section{Introduction}
Modern computing systems are rapidly increasing in complexity and scale to combat the slowdown of Dennard scaling and to satisfy the ever-increasing need for more computing performance and memory, driven by \gls{ml} and big data workloads~\cite{frazelle_chip_2021}. %
Kumar et al. highlight the importance of irregular memory accesses in sparse data structures when dealing with large-scale graph applications~\cite{kumar_efficient_2016}. %
Today's systems require high-performance interconnects with components that efficiently move the data required to supply their compute units. %
Using such specialized \glspl{dmac} is a well-established method to transfer data independently of processors, thereby promising to achieve high throughput while the processor is free to perform computationally useful work~\cite{su_processor-dma-based_2011, ma_design_2009, comisky_scalable_2000, chen_dma_2010, amd_xilinx_axi_2022, synopsys_designware_nodate}. %
With a shift towards more heterogeneous architectures, as well as smaller datatypes~\cite{frazelle_chip_2021, jang_encore_2022}, more diverse transfers are emerging, requiring greater flexibility and less overhead in the programmability of \glspl{dmac}. %

Highly flexible programming hardware interfaces and hardware abstraction layers for \glspl{dmac} are usually based on \emph{descriptors}, data structures stored in shared memory that hold the information of a transfer. %
Descriptors have multiple advantages compared to simpler register-based programming interfaces, which are widely used in embedded and \gls{mcu} applications~\cite{rossi_ultra-low-latency_2014}. %
Descriptors massively reduce the requirement for dedicated configuration memory space by storing the transfer specification in general-purpose memory segments, thus eliminating the need for configuration space replication in multicore applications to enforce atomic transfer launching~\cite{rossi_ultra-low-latency_2014}. %
Descriptors can be chained as a linked list, enabling the automatic launch of subsequent transfers. %
This allows multidimensional affine and fully arbitrary and irregular workloads to be processed~\cite{fjeldtvedt_cubedma_2019}. %
A concrete example of a \gls{dmac} is the \emph{LogiCORE IP DMA}, an \gls{axi} \gls{dma} soft \gls{ip} by \emph{Xilinx}~\cite{amd_xilinx_axi_2022}. It is a high-bandwidth \gls{dmac} with a descriptor-based programming interface and is designed to transfer data between a memory-mapped \gls{axi} interface and an \gls{axi}-Stream target device. %

Applying this descriptor configuration model to fine-grained, irregular transfers leads to long chains of individual descriptors, requiring the \gls{dmac} unit to handle a large amount of data when executing the transfer. %
Excessive descriptor size degrades the throughput of such fine-grained transfers, as the \gls{dmac} may require multiple cycles to fetch the descriptors. %
Furthermore, larger descriptor sizes result in more significant resource utilization and power overhead required by buffering logic. %
\emph{{Synopsys'}} \emph{DesignWare} \gls{axi} \gls{dma} controller presents a parametrizable and high-performance \gls{dmac} solution, using 64-byte-long descriptors in \emph{scatter-gather mode}~\cite{synopsys_designware_nodate}. %
Paraskevas et al. describe an efficient 32-byte-long descriptor format without prefetching capabilities~\cite{paraskevas_virtualized_2018}. %
In their work, descriptors are stored in dedicated pages of the core's on-chip scratchpad memory. %
Ma et al. describe a five-entry-long descriptor format supporting chaining for multidimensional data transfers~\cite{ma_efficient_2019}. %
To ensure efficient fetching of the descriptors, they are stored in a dedicated \gls{dmac}-internal \emph{parameter {RAM}}. %

As descriptors are usually handled in sequence~\cite{amd_xilinx_axi_2022}, requesting the next descriptor once the prior is read, full \gls{dmac} utilization is only reached if the described transfer is long enough to hide the latency of fetching a descriptor. %
This can no longer be guaranteed for fine-grained transfers in a non-ideal memory system, limiting the maximum achievable \gls{dmac} utilization for such transfers. %

In this work, we tackle both of these issues by introducing a \gls{dmac} with a minimal descriptor format, as well as a low-overhead prefetching mechanism; our contributions are:

\begin{enumerate}
    \item %
    A lightweight, minimal, and efficient descriptor format holding only the essential information required to describe a transfer. %
    Our format supports chaining and provides a mechanism to track transfer completion, increasing \gls{dmac} utilization by \todo{3.9$\times$} for \todo{64-byte} transfers  compared to the \emph{LogiCORE IP DMA}~\cite{amd_xilinx_axi_2022}.
    \item %
    Implementing our descriptor format, as well as speculative descriptor prefetching, together with an existing \gls{amba} \gls{axi} \gls{dma} engine~\cite{kurth_open-source_2022}, creating a fully parametrizable, synthesizable, and technology-independent \gls{dmac}. %
    \item %
    Evaluating our \gls{dmac} \gls{ooc} regarding its performance, area requirements, and timing in a 12~nm node. %
    The \gls{dmac} can achieve near-ideal performance while exceeding clock frequencies of \todo{1.44~GHz} and requiring only \todo{49.5~kGE}. %
    \item %
    Integrating the resulting \gls{dmac} into a 64-bit RISC-V SoC~\cite{zaruba_cost_2019} using general-purpose {DDR3} memory to store our descriptors. %
    We achieve an improvement in terms of latency by \todo{1.66$\times$} compared to the \emph{LogiCORE IP DMA} \gls{axi} \gls{dma}~\cite{amd_xilinx_axi_2022}, while requiring \todo{11\%} fewer lookup tables, \todo{23\%} fewer flip-flops, and no block {RAM}s.
\end{enumerate}

\section{Architecture}\label{sec:arch}

With efficient data transfer being an essential requirement for \gls{ml} workloads, we make use of the low-level \gls{dma} engine presented by Kurth et al. in~\cite{kurth_open-source_2022}. %
This fully open-source \gls{dma} engine directly interfaces with an \gls{amba} \gls{axi} memory system and is capable of asymptotically utilizing the available bandwidth. %
Furthermore, the \gls{dma} engine in~\cite{kurth_open-source_2022} is optimized for low area utilization, low transfer launch latencies, and high clock frequencies; however, it does not directly provide a programming interface to the system. %

In the following section, we describe a descriptor-based programming interface called \emph{\gls{dma} frontend}. %
The \emph{\gls{dma} frontend} and \emph{\gls{dma} engine} as \emph{backend} together form the \gls{dmac}, as shown in \Cref{fig:frontend_overview}. %

\subsection{DMA Frontend Design}\label{sec:arch:dma-ctrl}

\begin{figure}[t]
    \centering
    \includegraphics[width=\columnwidth]{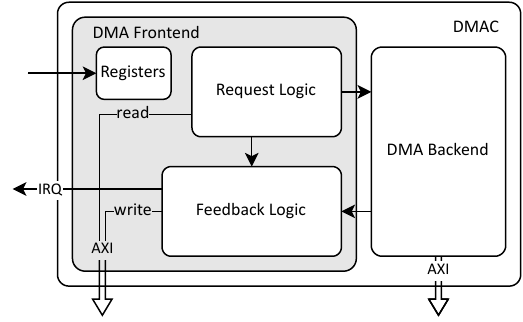}
    \caption{Overview of the \gls{dmac}, containing request logic with internal registers for configuration and read logic to fetch the descriptor, and feedback logic to update the system once the \gls{dma} backend completes the transfer.}
    \label{fig:frontend_overview}
\end{figure}

To configure a \gls{dma} transfer, the \gls{dma} frontend exposes a memory-mapped \gls{csr}, which accepts an address pointing to a \gls{dma} transfer descriptor in shared memory, described in \Cref{sec:arch:descr}. %
Once this pointer is written into the \gls{csr}, the frontend requests the descriptor from memory through the read channel of an \gls{axi} manager port, shown as the request logic in \Cref{fig:frontend_overview}. %
This manager port is configurable in both \gls{axi} address width, allowing descriptors to be located in any memory location, and \gls{axi} data width, ranging from \SIrange{16}{512}{bits}. %
Using this port, the frontend retrieves the necessary information for a generic linear memory transfer: source address pointer, destination address pointer, transfer length, and configuration. %

Once fetched, the frontend forwards the information to the \gls{dma} backend, which executes the transfer. %
To improve performance, both the \gls{csr} and the connection to the backend implement a queue. %
This allows multiple transfers to be enqueued, maximizing the utilization of the backend.

Once the \gls{dma} backend has completed a transfer, the frontend reports completion back to the system, shown as the feedback logic in \Cref{fig:frontend_overview}. %
For each transfer, the corresponding descriptor is modified to indicate its completion, and an interrupt is signaled if configured. %

\subsection{\gls{dmac} Transfer Descriptor}\label{sec:arch:descr}

Our \gls{dmac} descriptor contains the information necessary to fully describe a linear memory transfer: a 64-bit source and destination address, and the length of the transfer. %
The transfer length is stored in an unsigned 32-bit field, allowing individual transfers of up to 4~GiB in size. %
Longer transfers can easily be achieved by chaining together multiple descriptors. %

A \emph{config} field in the descriptor holds configuration information for both the \gls{dma} front- and backend. %
For the former, different \gls{irq} options can be set, while for the latter, various \gls{axi}-related parameters are configurable. %
A complete descriptor structure can be found in \Cref{lst:descriptor}. %

Apart from information describing the transfer, the descriptor contains a pointer to the next descriptor to be processed, enabling \emph{descriptor chaining}. %
This allows our \gls{dmac} to process a \emph{linked list} or \emph{chain of descriptors} in memory without involving the \gls{cpu}. %
The last descriptor in a chain carries all ones (equals to -1) in the \emph{next} field; we call this value \emph{end-of-chain}. This value was chosen as no descriptor can fit at the corresponding address. %
Descriptor chaining allows the construction of arbitrary and irregular transfers from simple linear transfers. %

When designing the descriptor format, we minimized its size while keeping it a multiple of the \gls{axi} bus width. %
The former has two benefits; it not only reduces the required bandwidth of the memory subsystem when storing and fetching descriptors, but also the overall memory footprint to describe a given transfer. %
The latter allows us to fetch the 256-bit descriptors and chains thereof without losing utilization in memory systems with widths up to 256~bits. %
In systems featuring a 512-bit infrastructure, such as a wide range of \emph{Xilinx Zynq UltraScale+ MPSoCs}, two full descriptors could be fetched in one cycle. %

\begin{lstlisting}[language=C, caption={Descriptor Layout}, label={lst:descriptor}]
struct descriptor {
  u32 length;
  u32 config;
  u64 next;
  u64 source;
  u64 destination;
}
\end{lstlisting}

To compare, the \emph{LogiCORE IP DMA}~\cite{amd_xilinx_axi_2022} uses a descriptor format of thirteen 32-bit words or 416~bits, of which usually only 256~bits are read. %
Its \gls{axi} manager interface used to fetch descriptors is limited to a data width of 32~bits, leading to a descriptor read latency of at least eight to thirteen cycles. %
In contrast, our \gls{dmac} may read a descriptor in four cycles in a comparable 64-bit system. %

\subsection{Speculative Descriptor Prefetching}\label{sec:arch:prefetch}

To compensate for memory latency, we employ \emph{speculative descriptor prefetching}. %
Once a descriptor address is written to the \gls{csr}, we not only request the first descriptor over the frontend's manager interface but send up to a configurable amount of requests with sequential addresses. %
The number of descriptors speculatively requested is configured using the \emph{prefetching} compile-time parameter, zero deactivating the prefetching logic, as can be seen in \Cref{tab:param}. %

Once a descriptor arrives at the \gls{dma} frontend, we compare the \emph{next} field of this descriptor with the speculatively requested address. %
On a match, the speculative address is committed and one speculation slot is freed up. %
Should a misprediction occur, we discard all descriptor addresses in the \emph{speculation slots} and start to fetch from the correct \emph{next} address while ignoring the incoming data that was mispredicted. %

Care was taken not to introduce any latency in the case of mispredictions: Assuming there is space in the \emph{speculation slots}, the proper request is issued over the \gls{axi} manager interface in the same cycle the \gls{dma} frontend receives the \emph{next} field. %
This is the same latency we observe in the case of prefetching disabled. %
Thus, the only performance degradation that may occur is caused by \todo{minimal} additional contention in the memory system due to fetching data that is directly discarded. %

\subsection{SoC integration}\label{sec:arch:soc}

\begin{figure}
    \centering
    \includegraphics[width=\columnwidth]{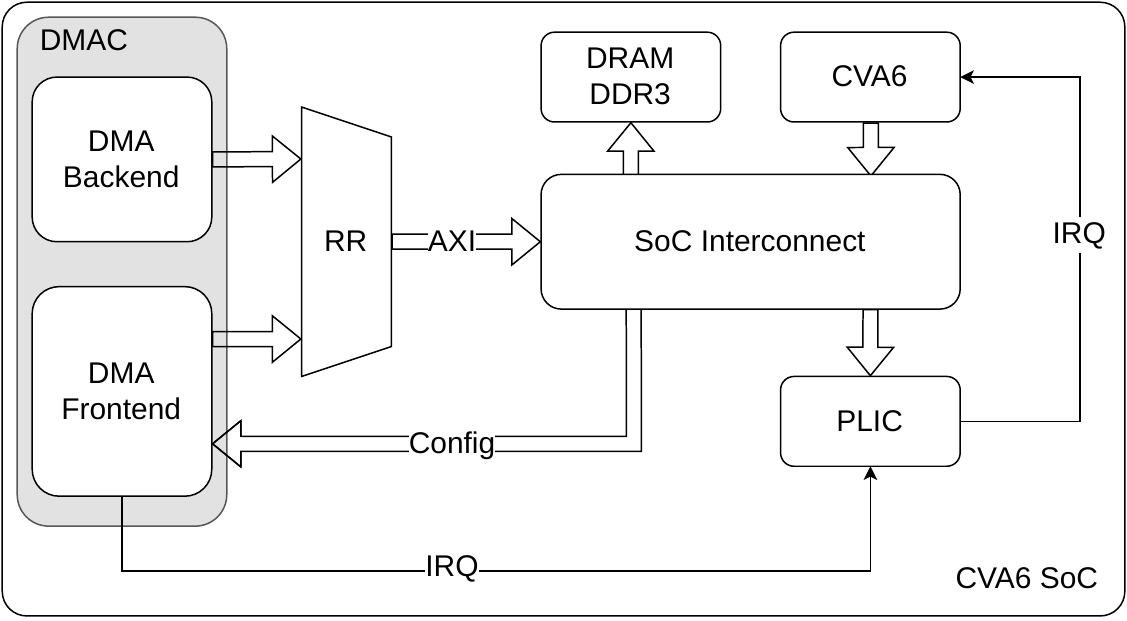}
    \caption{Integration of our \gls{dmac} into the CVA6 \gls{soc}. The two manager interfaces, after arbitration,  as well as the subordinate configuration port of the \gls{dmac}, are connected to the \gls{soc}'s interconnect. The \gls{irq} line is connected to the platform's \gls{plic}. }
    \label{fig:integration_soc}
\end{figure}

Heterogeneous systems often rely on a high-performance 64-bit memory system due to compatibility with the central host processor. %
While parametrizable, our implementation configures the \gls{dmac} for such a memory system, using 64~bits both for address and data width of the \gls{axi} bus in accordance with the \emph{CVA6 RISC-V SoC}~\cite{zaruba_cost_2019} \todo{we integrate our \gls{dmac} into.} %
An overview of the resulting system can be seen in \Cref{fig:integration_soc}: The two manager interfaces of our \gls{dmac}, as well as the subordinate configuration interface, are connected to the memory system of the \gls{soc}. %

We occupy one new \gls{irq} channel at the systems' \gls{plic}, which is used to signal transfer completion when configured. %
For lightweight in-system progress reporting, we repurpose a transfer descriptor by overwriting its first 8 bytes with \emph{all ones} after the transfer is completed. %
This allows us to forego raising an interrupt after each linear transfer is completed, thus making interrupt notification optional. %

\subsection{Linux Driver}

To ensure simple integration into existing environments, we provide a sample \emph{Linux} driver with an accompanying device-tree file. %
The \gls{dma} subsystem of the Linux kernel exposes a broad \gls{api}~\cite{noauthor_linux_nodate}, of which we implement the \emph{memcpy} interface. %

For a \gls{dma} client to request a data transfer, the application requests the driver to prepare the \emph{memcpy} transfer. %
This is done by allocating one or more chained descriptors and populating \emph{source}, \emph{destination}, \emph{length}, and \emph{config} fields. %
Should a transfer consist of more than one descriptor, then only the last has \gls{irq} signaling enabled.

As a second step, the client commits to specific transfers, which results in the driver chaining them in a FIFO fashion to a new chain. %

Third, the client requests to submit all committed transfers to the hardware. %
The driver checks whether less than the maximum number of allowed chains are already running on the \gls{dmac}; if so, it schedules the new chain with a write to the \gls{dmac}'s \gls{csr}, otherwise, the transfers are stored to be scheduled later. %

Finally, on transfer completion, the \gls{dmac} raises an \gls{irq}. %
This leads to a call of the \emph{interrupt handler}, which schedules any completion callbacks the client has registered, updates the number of active chains if the transfer was the last of a chain, and schedules stored transfers. %

\section{Results}\label{sec:results}

We evaluate our \gls{dmac} with our optimized descriptor format \acrfull{ooc} and in-system. %
For the \gls{ooc} evaluation, we attach our \gls{dmac} to a configurable memory system to assess its performance, as can be seen in \Cref{fig:ooc-testbench}. %
We then present area and timing results from synthesizing our controller out-of-context using a 12~nm {FinFET} node. %

We then show both performance and implementation results of our \gls{dmac} integrated into a 64-bit RISC-V CVA6 \gls{soc}~\cite{zaruba_cost_2019} emulated on a \emph{Diligent Genesys 2} \gls{fpga}~\cite{noauthor_genesys_nodate}. %

\subsection{Out-of-context Results}\label{sec:ooc-results}

\begin{figure}
    \centering
    \includegraphics[width=\columnwidth]{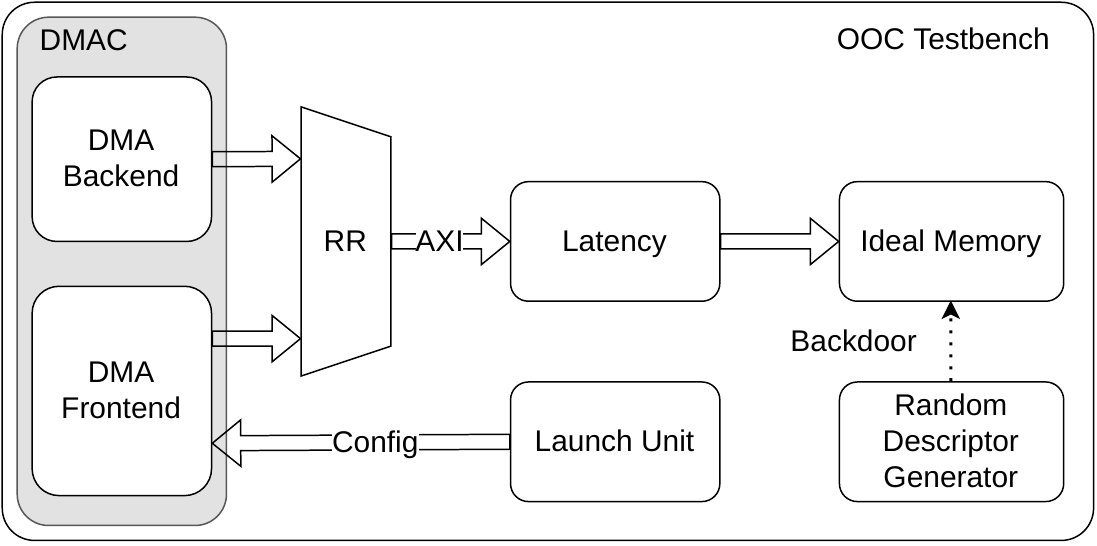}
    \caption{The \gls{ooc} testbench setup; the \gls{dmac} has its two \gls{axi} manager interfaces connected to a fair round-robin arbiter (RR), which in turn is connected to a latency-configurable memory system. Descriptors are loaded into the memory using backdoor access and are launched via the \gls{dmac}'s subordinate configuration interface.}
    \label{fig:ooc-testbench}
\end{figure}

To evaluate the standalone performance of our \gls{dmac}, we created a testbench environment consisting of a \emph{latency-configurable} memory system and a \todo{\emph{launch unit}} to set up and execute random streams of descriptors. %
To simulate a real system, both of our \gls{dmac}'s \gls{axi} manager ports are connected to the same memory system using a fair round-robin arbiter ({RR}), as shown in \Cref{fig:ooc-testbench}. %
To stay aligned with our target CVA6 \gls{soc}, we set the address and data width of our \gls{ooc} testbench to 64~bits. %

The randomness of the descriptions can be closely controlled, allowing us to emulate different transfer characteristics. %
The corresponding descriptors are immediately preloaded into our simulation memory using a backdoor, while the actual launch of the transfers is controlled using our \gls{dmac}'s \gls{csr} interface. %

The bus utilization is measured at the \gls{dma} backend's \gls{axi} \emph{manager} interface; only \textit{useful} payload traffic contributes to utilization. %
We only report \emph{steady state} bus utilization suppressing any possible cold-start phenomena. %

In our analysis, we assessed three distinct memory system configurations reflecting different use cases:
\begin{enumerate}
    \item \emph{Ideal Memory:} We configure our simulation memory to have one cycle latency emulating an SRAM-based main memory. %
    \item \emph{DDR3 Main Memory:} Replicating the conditions found on the \emph{Diligent Genesys 2} \gls{fpga}~\cite{noauthor_genesys_nodate} when accessing DDR3 off-chip memory, we include a configuration with \textit{thirteen} cycles latency. %
    \item \emph{Ultra-deep Memory:} Representing a large \gls{noc} system found in a modern \gls{soc}, we include a configuration with a latency of \textit{one hundred} cycles. %
\end{enumerate}

\begin{table}
    \caption{The compile-time parameters used in our \gls{ooc} experiments.}
    \begin{center}
    \renewcommand{\arraystretch}{1.1} %
    \resizebox{0.9\columnwidth}{!}{
    \begin{tabular}{ccc}
    \hline
    \textbf{Configuration} & Descriptors In-flight & Prefetching \\
    \hline
    \emph{LogiCORE IP DMA}~\cite{amd_xilinx_axi_2022} & 4 & \textit{N.A.} \\
    \emph{base} & 4 & Disabled (0) \\
    \emph{speculation} & 4 & 4 \\
    \emph{scaled} & 24 & 24 \\
    \hline
    \end{tabular}
    }
    \label{tab:param}
    \end{center}
    \vspace{\tablermws}
\end{table}

To ensure a fair comparison against the \emph{LogiCORE IP DMA}, we include a \emph{base} configuration closely matching the \emph{LogiCORE IP DMA}'s default configuration. %
In our evaluation, we included two additional configurations; one enabling \emph{speculation} while closely resembling the \emph{base} configuration and a \emph{scaled} configuration setting both the \todo{number of \emph{descriptors in-flight}} and the \emph{prefetching} to \textit{24}. %
We summarize the respective parameter configurations in \Cref{tab:param}. %

\begin{table*}
    \caption{Area requirements at the maximum clock frequency of the \gls{dmac} and its main sub-components; the \gls{dma} frontend and the \gls{dma} backend. Clock frequencies are achieved in typical conditions.}
    \begin{center}
    \renewcommand{\arraystretch}{1.1} %
    \resizebox{0.8\linewidth}{!}{
    \begin{tabular}{cccc|c}
    \hline
    \textbf{Configuration} & \textbf{DMA Frontend} & \textbf{DMA Backend} & \textbf{Total DMAC} & \textbf{Achievable Clock Frequency} \\
    \hline
    \emph{base}        & 25.8~kGE  & 15.4~kGE & 41.2~kGE  & 1.71~GHz \\
    \emph{speculation} & 34.8~kGE  & 14.7~kGE & 49.5~kGE  & 1.44~GHz \\
    \emph{scaled}      & 151.1~kGE & 37.3~kGE & 188.4~kGE & 1.23~GHz \\
    \hline
    \end{tabular}
    }
    \label{tab:area_ooc}
    \end{center}
\end{table*}

\begin{figure}
    \centering
    \label{fig:ooc}
    \subfloat[Ideal memory \textit({1 cycles latency)}]{%
        \includegraphics[width=0.95\columnwidth]{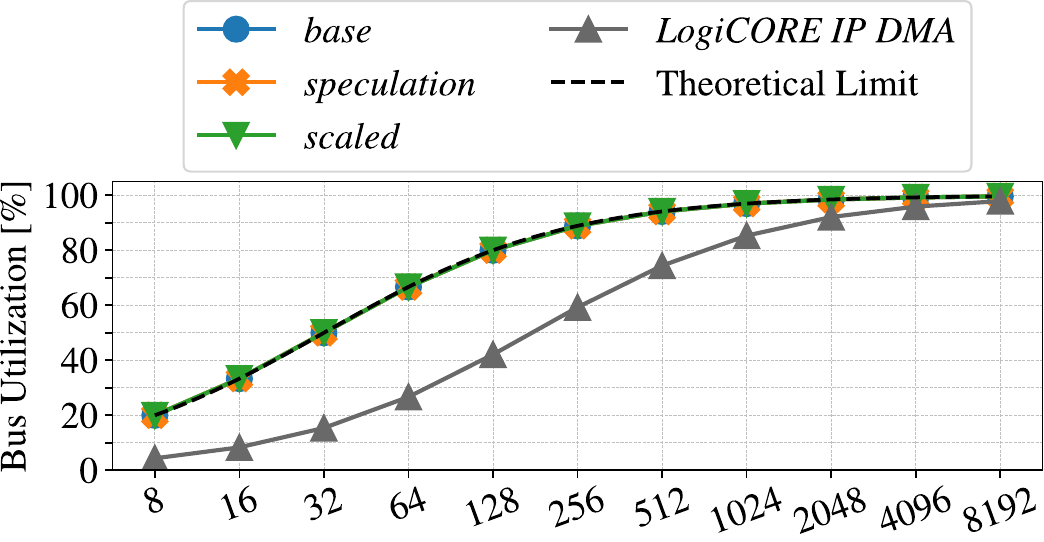} \label{fig:ooc:0}
    }

    \subfloat[DDR3 main memory \textit({13 cycles latency)}]{%
        \includegraphics[width=0.95\columnwidth]{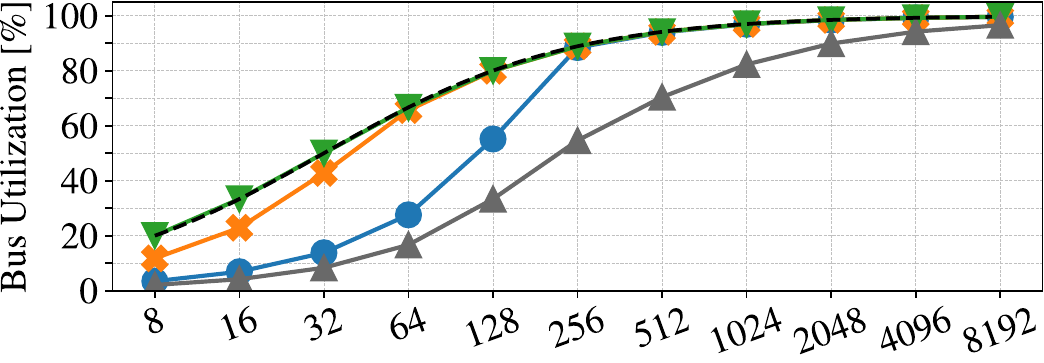} \label{fig:ooc:13}
    }

    \subfloat[Ultra-deep memory \textit({100 cycles latency)}]{%
        \includegraphics[width=0.95\columnwidth]{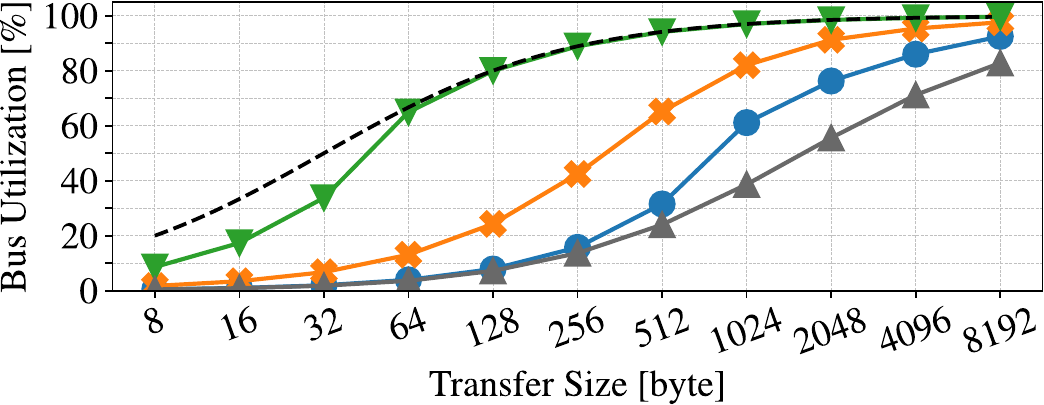} \label{fig:ooc:100}
    }
    \caption{\Gls{dmac} \emph{steady-state} bus utilization given a prefetch hit rate of 100\% in memory systems featuring various latencies.}
\end{figure}

\begin{figure}
    \centering
    \includegraphics[width=0.95\columnwidth]{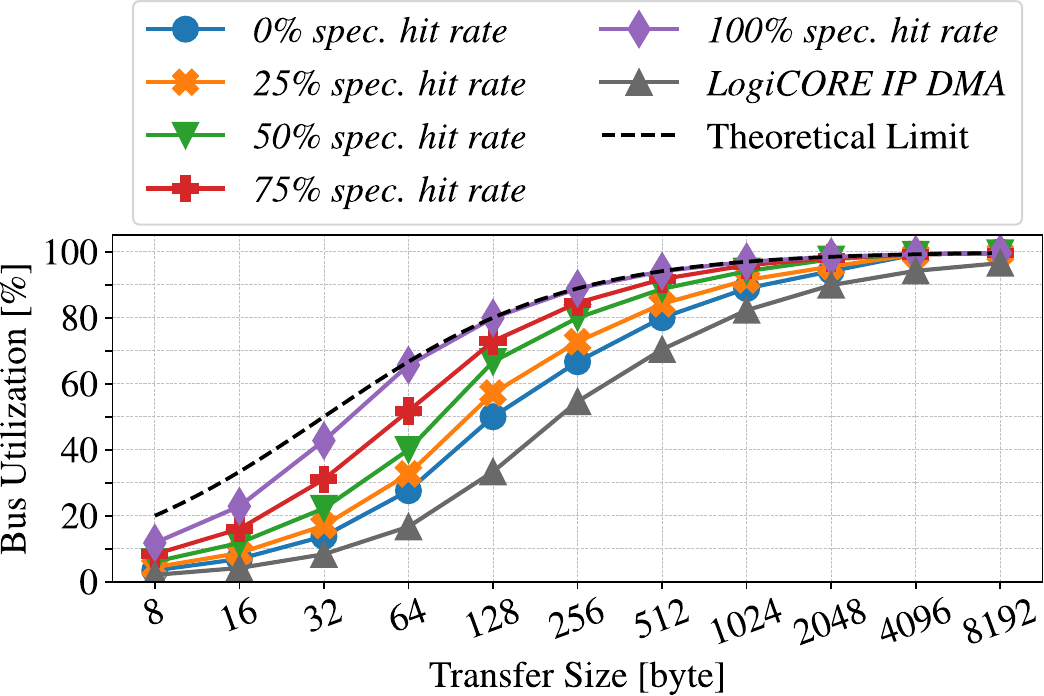}
    \caption{\Gls{dmac} steady-state bus utilization in the case of the DDR3 main memory with speculation misses; \emph{speculation} configuration.}
    \label{fig:cache_misses}
\end{figure}

As access to main memory is shared between the \gls{dma} frontend and \gls{dma} backend, the bus utilization, as defined above, cannot reach 100\%. %
The transfer of the payload will be interrupted by descriptor transfers, limiting the \emph{ideal bus utilization}, $\bar{u}$ -- see \Cref{eq:ideal-util} -- where $n$ is the transfer size in byte. %

\begin{equation}\label{eq:ideal-util}
\bar{u} = \frac{n}{n + 32}
\end{equation}

Descriptor misprediction, in the case of speculative prefetching enabled, further limits the ideal utilization, as it inflates the number of additional bytes fetched by the \gls{dma} frontend per transfer. %

In very shallow or ideal memory systems, our \emph{base} configuration already achieves ideal steady-state utilization for any bus-aligned transfer size, as shown in~\Cref{fig:ooc:0}. %
At transfer sizes of 64~B -- a typical cache line size in many memory architectures -- we improve the utilization by \todo{2.5$\times$} compared to the \emph{LogiCORE IP DMA}. %

When using the \emph{Genesys 2 DDR3} latency configuration, we achieve ideal steady-state utilization at \todo{256~B} without and \todo{64~B} with prefetching enabled, as can be seen in~\Cref{fig:ooc:13}. %
This increases the utilization by up to \todo{1.7$\times$} and \todo{3.9$\times$}, respectively, compared to the \emph{LogiCORE IP DMA}. %

Finally, we show that our \gls{dmac} can be configured to still achieve near-ideal steady state utilization even in ultra-deep memory systems. As can be seen in \Cref{fig:ooc:100}, the \emph{scaled} configuration achieves ideal utilization starting from \todo{128~B}.

Varying prefetching hit rates of \SIrange{75}{0}{\%}, our achievable increase in bus utilization compared to the \emph{LogiCORE IP DMA} still ranges from \SIrange{1.65}{3.1}{\times} at \todo{64~B}, see \Cref{fig:cache_misses}. %

We evaluate the timing and resource requirements of our \gls{dmac} in the various configurations presented in \Cref{tab:param} by synthesizing our work in \emph{{GlobalFoundries'} {GF12LP+} FinFET} technology using \emph{Synopsys' Design Compiler NXT} in \emph{topological} mode. %
All results are presented in the typical corner of the library at \todo{25\textcelsius} at \todo{0.8~V}, in \Cref{tab:area_ooc}. %

Our \emph{base} configuration requires an area of 41.2~kGE, achieving a maximum clock frequency of 1.71~GHz. %
Enabling prefetching adds 8.3~kGE while reducing the achievable maximum clock frequency to 1.44~GHz. %

We synthesized our design in numerous configurations, creating a model of the circuit area as a function of the parameters in \Cref{tab:param}. %
The design's area in \textit{kGE} is described by: $A = 20.30 + 5.28 \* d + 1.94 \* s$, where $d$ denotes the number of descriptors in flight and $s$ the number of speculatively launched descriptors. %
The total area is linear in $d$ and $s$, allowing the hardware to be easily scaled to larger sizes. %

The \emph{scaled} configuration requires a total of 188.4~kGE achieving 1.23~GHz. %
Comparing these numbers to CVA, we find the \gls{dmac} area to be less than \todo{10~\%} of the core's area while achieving  similar clock speeds, confirming the scalability of our controller.

\subsection{In-system Results}\label{sec:soc-results}

To evaluate the required resources on a \gls{fpga}, we synthesized the CVA6-\gls{soc}~\cite{zaruba_cost_2019} with the various configurations of our \gls{dmac} integrated. %
Synthesis was done using \emph{Vivado 2019.2} targeting the \emph{Genesys 2} board, which features a \emph{Kintex 7} \gls{fpga} from \emph{Xilinx}. %

In the \emph{base} configuration, the footprint of the \gls{dmac} is \todo{2610} \glspl{lut} and \todo{3090} \glspl{ff}, while the entire \gls{soc} occupies \todo{79142} \glspl{lut} and \todo{58086} \glspl{ff}, see \Cref{tab:util_soc_fpga}. %
This puts the base configuration at \todo{3.3\%} of total \gls{lut} usage and \todo{5.3\%} of total \gls{ff} usage, and is a reduction of \todo{6.25\%} \gls{lut} and \todo{39.8\%} \gls{ff} utilization compared to the \emph{LogiCORE IP DMA}. %

Compared to the \emph{base} configuration, the \emph{speculation} configuration uses \todo{27\%} more \glspl{ff}, but reduces the number of \glspl{lut} by \todo{5\%}. %
The scaled configuration increases resource utilization further, requiring \todo{2.59$\times$} as many \glspl{lut} and \todo{3.67$\times$} as many \glspl{ff} as the \emph{base} configuration. %

\begin{table}
    \caption{\gls{fpga} resource requirements of the \gls{dmac} at 200~MHz.}
    \begin{center}
    \renewcommand{\arraystretch}{1.1} %
    \begin{tabular}{ccc}
    \hline
    \textbf{Configuration} & \textbf{LUTs} & \textbf{FFs} \\
    \hline
    \emph{base}          & 2610 & 3090 \\
    \emph{speculation}   & 2480 & 3935 \\
    \emph{scaled}        & 6764 & 11353 \\
    \emph{LogiCORE IP DMA}~\cite{amd_xilinx_axi_2022} & \textit{2784} & \textit{5133} \\
    \hline
    \end{tabular}
    \label{tab:util_soc_fpga}
    \end{center}
    \vspace{\tablermws}
\end{table}

We use our latency-configurable memory system presented in \Cref{sec:ooc-results}, which we integrate into the upstream CVA6-\gls{soc} to measure the following three different latencies:
\begin{itemize}
    \item \texttt{i-rf:} the \gls{cpu} issuing a write to the \gls{dma} frontend issuing a read request %
    \item \texttt{rf-rb:} between the issue of the read request from \gls{dma} frontend and the \gls{dma} backend %
    \item \texttt{r-w:} the latency between the \gls{dma} engine reading and writing the same data %
\end{itemize}

\begin{table}
    \caption{\gls{dmac} latencies between various events and memory systems for the \emph{scaled} configuration}
    \begin{center}
    \renewcommand{\arraystretch}{1.1} %
    \begin{tabular}{lccccc}
        \hline
        \textbf{Metric} && \emph{LogiCORE IP DMA} \gls{dma}~\cite{amd_xilinx_axi_2022} & \emph{scaled} \\
        \hline
        \texttt{i-rf} && 10  & 3 \\
        \texttt{rf-rb} &
        1 cycle latency & 22 & 8 \\
        &13 cycles latency & 48 & 32 \\
        &100 cycles latency & 222 & 206 \\
        \texttt{r-w}  && 1 & 1  \\
        \hline
    \end{tabular}
    \label{tab:latencies}
    \end{center}
    \vspace{\tablermws}
\end{table}

As can be seen in \Cref{tab:latencies}, we achieve \todo{three} cycles of latency for \texttt{i-rf}, an improvement of \todo{3.33$\times$} over the \emph{LogiCORE IP DMA}. %
For \texttt{rf-rb}, we achieve a latency of \todo{eight} cycles in ideal memory, \todo{32} cycles with a memory latency of 13, and \todo{206} cycles in the case of 100 cycles of latency. %
This results in an improvement of \todo{2.75$\times$}, \todo{1.5$\times$}, and \todo{1.08$\times$}, respectively. %
Latencies for \texttt{r-w} are equal at \todo{one} cycle for both our \gls{dmac} and the \emph{LogiCORE IP DMA}. %

\section{Conclusion}

In this work, we present a scalable, platform-independent, synthesizable \gls{dmac} for fast and efficient data transfers in \gls{axi}-based systems. %
Compared to a competing solution, we achieve \todo{1.66$\times$} less latency, increasing bus utilization by up to \todo{2.5$\times$} in an ideal memory system with \todo{64-byte transfers}, overall requiring \todo{11\%} fewer \glspl{lut} and \todo{23\%} fewer \glspl{ff} without requiring any block {RAM}s. %
In deep memory systems, we show an even more significant increase in the utilization of \todo{3.6$\times$} with \todo{64-byte transfers}. %

We show the utility of our \gls{dmac} in a 64-bit Linux-capable environment by implementing it into an \gls{soc} based around {CVA6} and present a working driver for Linux. %
The resulting \gls{dmac} is available fully open-source.\footnote{\wheretofind} %

%

\input{main.bbl}

%
\end{document}

%% file: main.bbl
% Generated by IEEEtran.bst, version: 1.14 (2015/08/26)